\documentclass[aps,prb,twocolumn,showpacs]{revtex4}
\usepackage{bm}
\usepackage{graphicx}
\usepackage{amsmath}
\usepackage{eufrak}
\usepackage{color}
\usepackage{upgreek}
\usepackage{subfigure}
\usepackage{hyperref}
\newcommand{\nix}[1]{}

\begin{document}

\title{Cyclotron Resonance Assisted Photocurrents in Surface States of a 
3D Topological Insulator Based on a Strained High Mobility HgTe Film
}

\author{K.-M.\,Dantscher,$^1$  D.A.\,Kozlov,$^{2,3}$ P.\,Olbrich,$^1$ C.\,Zoth,$^1$ 
P.\,Faltermeier,$^1$ M.\,Lindner,$^1$ G.V.\,Budkin,$^4$ S.A.\,Tarasenko,$^{4,5}$ 
V.V.\,Belkov,$^4$ Z.D.\,Kvon,$^{2,3}$ N.N.\,Mikhailov,$^2$ S.A.\,Dvoretsky,$^2$  D.\,Weiss,$^1$ 
B. Jenichen,$^6$  
and S.D.\,Ganichev$^1$}
\affiliation{$^1$Terahertz Center, University of 
Regensburg, 93040 Regensburg, Germany}

\affiliation{$^2$A.V. Rzhanov Institute of
Semiconductor Physics, Novosibirsk 630090, Russia}
\affiliation{$^3$Novosibirsk State University,
Novosibirsk 630090, Russia}

\affiliation{$^4$Ioffe Physical-Technical Institute, 
194021 St.\,Petersburg, Russia}

\affiliation{  $^5$St.\,Petersburg State Polytechnic University, 195251 St.\,Petersburg, Russia }

\affiliation{$^6$ Paul-Drude-Institut for Solid State Electronics, 
10117 Berlin, Germany}

\begin{abstract}
We report on the observation of cyclotron resonance induced photocurrents, 
excited by continuous wave terahertz radiation, in a 3D topological insulator (TI) based on an 80~nm strained HgTe film. The analysis of the photocurrent formation 
is supported by complimentary measurements of magneto-transport and radiation transmission. 
We demonstrate that the photocurrent is generated in the topologically protected surface states. 
Studying the resonance response in a gated sample we examined the behavior of the photocurrent, 
which enables us to extract the mobility and the cyclotron mass as a function of the Fermi energy.
For high gate voltages we also detected cyclotron resonance (CR) of bulk carriers, with a mass about two times larger than that obtained for the surface states. The origin of the CR assisted photocurrent 
is discussed in terms of asymmetric scattering of TI surface carriers in the momentum space.
Furthermore, we show that studying the photocurrent 
in gated samples provides a sensitive method to probe the effective masses and the mobility of 2D Dirac surface states, when the Fermi 
level lies in the bulk energy gap or even in the conduction band.
\end{abstract}

\maketitle

\section{Introduction}

The physics of relativistic Dirac fermions in semiconductors has recently moved 
into the focus of modern research, due to their unique properties and promising 
novel applications, for recent reviews 
see~\cite{CastroNeto2009,Novoselov2012,Glazov2014,Moore2010,
Hasan2010,Zhang2011}. 
Among diverse systems addressed in the literature HgTe-based structures represent 
an extraordinary material class, allowing the fabrication of a high quality material~\cite{Becker2003,Becker2003_2,Dvoretsky2010,Baenninger2012,Ballet2014,Becker2014} with Dirac-like 
systems of different forms. The latter includes topological protected edge and 
surface states of two- and three-dimensional topological insulators 
\cite{Bernevig2006,Koenig2007,Kane2007,Dai,Roth2009,Bruene2011,Hancock2011,Nowack2013,Grabecki2013,Oostinga2013,Kvon2014,Kozlov2014,Bruene2014,Chen2014}, QWs with critical 
thickness \cite{Bernevig2006,Buettner2011,Kvon2011,Zholudev2012,Olbrich2013,Ludwig2014,Zoth2014,Pakmehr2014,Tarasenko2015} 
and bulk HgCdTe material at the point of semiconductor-to-semimetal transition \cite{Orlita}. Three 
dimensional TIs based on strained HgTe films \cite{Bruene2011} are of 
particular interest. Indeed, in these materials strain opens a gap in 
the otherwise gapless HgTe, which together with the high quality of the
 material allows one to obtain insulation in the bulk and to study 
electron transport in surface states only \cite{Bruene2011,Kozlov2014}. 
This differs significantly from all other known 3D TI (e.g. Bi$_2$Te$_3$, Sb$_2$Te$_3$), where $dc$ 
electron transport is (almost always) hindered by the high residual carrier 
density in the bulk\cite{bulk_states1,bulk_states2,bulk_states3}. This 
unique property allows to observe of the quantum Hall effect \cite{Bruene2011} 
and Shubnikov-de Haas oscillations \cite{Kozlov2014} in strained HgTe film and, thus to analyse surface state transport in TIs. Furthermore, the negligible contribution from bulk carriers opens 
a way to study Dirac fermions in 3D TI by cyclotron resonance measured via 
transmission or Faraday effect \cite{Shuvaev2012,Shuvaev,Shuvaev2013} and, as we show below, 
terahertz (THz) radiation induced photocurrents.

Here, we report on the observation of cyclotron resonance assisted  
photocurrents, excited in surface states of strained HgTe films. The analysis 
of the  photocurrent formation in topologically protected surface states is 
supported by complimentary measurements of the radiation transmission 
and magneto-transport. We demonstrate the photocurrent stems from magnetic field induced asymmetric scattering of 
nonequilibrium surface carriers in the momentum space. 
Our results reveal that studying the photocurrent 
provides a sensitive method to probe the effective masses of 2D Dirac surface 
states. Importantly, the method is applicable even for small size samples,
e.g. gated structures, and for Fermi level lying in the conduction band,
i.e.\,under conditions where transmission experiments are almost impossible.

\begin{figure}[t]
\includegraphics[width=\linewidth]{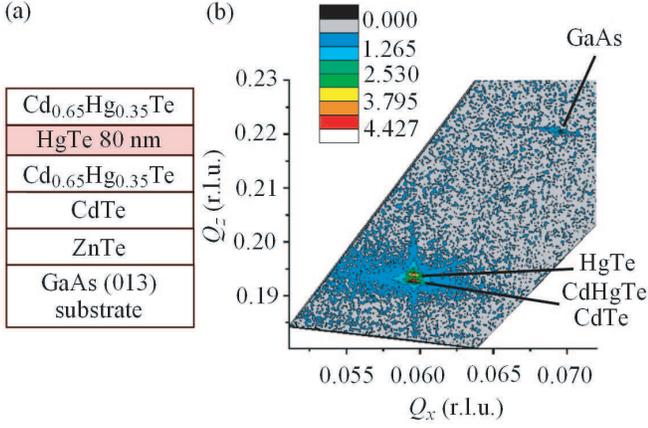}
\caption{(a) Cross section of the investigated structures. (b) 
A $X$-ray reciprocal space map near the asymmetric (113) reflection. 
The diffracted intensity is plotted on a color coded logarithmic scale.
 The CdTe film is fully relaxed with respect to the GaAs substrate, because 
the line connecting their maxima points to the origin of the reciprocal space. 
The HgTe and CdHgTe films are fully strained with respect to the CdTe, as their 
peaks lie on a straight line along the $z$-axis perpendicular to the sample 
surface. (r.\,l.\,u\,-\,reciprocal lattice unit)} 
\label{fig_1}
\end{figure}

\section{Samples}

The experiments are carried out on  molecular beam epitaxy grown high-mobility 80\,nm thick HgTe films. 
The HgTe film is sandwiched between thin Cd$_{0.65}$Hg$_{0.35}$Te layers acting as capping 
(top) and buffer (bottom) layers. 
The cross section of the structure is shown in Fig.\,\ref{fig_1}(a).  
The structure is grown on a GaAs substrate with (013) surface orientation
covered by a thin ZnTe layer and 4\,$\mu$m thick CdTe layer.  
The CdTe layer is fully relaxed because of its large thickness. 
The CdHgTe/HgTe/CdHgTe layers grown on top adopt the underlying CdTe lattice constant. The lattice mismatch between HgTe and CdTe of about 0.3\%, 
results in an tensile strain in the HgTe film, which opens a topological gap~\cite{Bruene2011}. 
The evidence for substantial strain in our samples comes from $X$-ray diffraction measurements, shown in 
Fig.\,\ref{fig_1}(b). Note, that such a sandwich design of the structure reduces 
the influence of dislocations caused by the lattice mismatch\cite{Kozlov2014} and 
allows one to obtain  high  electron mobilities  (up to 4$\times$10$^{5}$\,cm$^2$/V$\cdot$s) 
together with low residual bulk impurity concentration 
(about 10$^{16}$\,cm$^{-3}$ as estimated for ungated samples).

\begin{figure}[t]
\includegraphics[width=\linewidth]{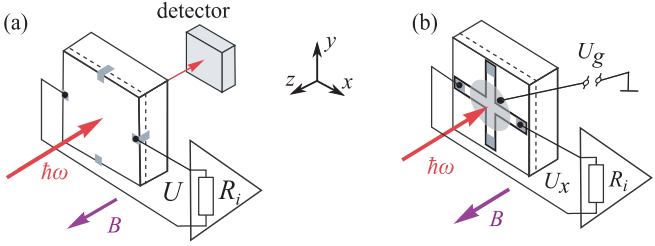}
\caption{Experimental setup of (a) the photocurrent and radiation transmission measurements in squared shaped samples and (b) photocurrent 
measurements in cross shaped samples.}
\label{fig_2}
\end{figure}

Several kinds of gated and ungated samples
have been prepared  from the same wafer, including ungated square shaped 
samples of $4\,$x$\,5$\,mm size, see 
Fig.\,\ref{fig_2}(a),  gated (ungated) cross shaped structures, 
see Fig.\,\ref{fig_2}(b), and gated hallbar samples. On the square shaped samples
we have fabricated eight ohmic contacts in the middle of the edges. The sample edges are grown along $x \parallel [100]$ and $y \parallel [0 3 \overline{1}]$. The samples of this kind have been used for simultaneous transmission and photocurrent measurements. In order to study photocurrents and magneto-transport
as a function of the  Fermi level position we have fabricated gated structures. 
To avoid insulator leakages we have used small area cross shaped structures  with four ohmic contacts,  Fig.\,\ref{fig_2}(b).
The structures of $50\,\upmu$m width and $1500\, \upmu$m length have been patterned 
by means of standard photolithography and wet etching.
Semitransparent Ti/Au gates of $20$\,nm/$5$\,nm thickness and $1500$\,$\upmu$m diameter have been 
deposited on $100$\,nm SiO$_{2}$ and $200$\,nm of Si$_{3}$N$_{4}$ dielectric layer grown by 
chemical vapour deposition. To monitor the influence of the gates on the CR position 
we have used both, gated as well as ungated, cross structures.
Gates have also been deposited on hallbar samples, which have been used primarily for
magneto-transport experiments.

\begin{figure}[t]
\includegraphics[width=\linewidth]{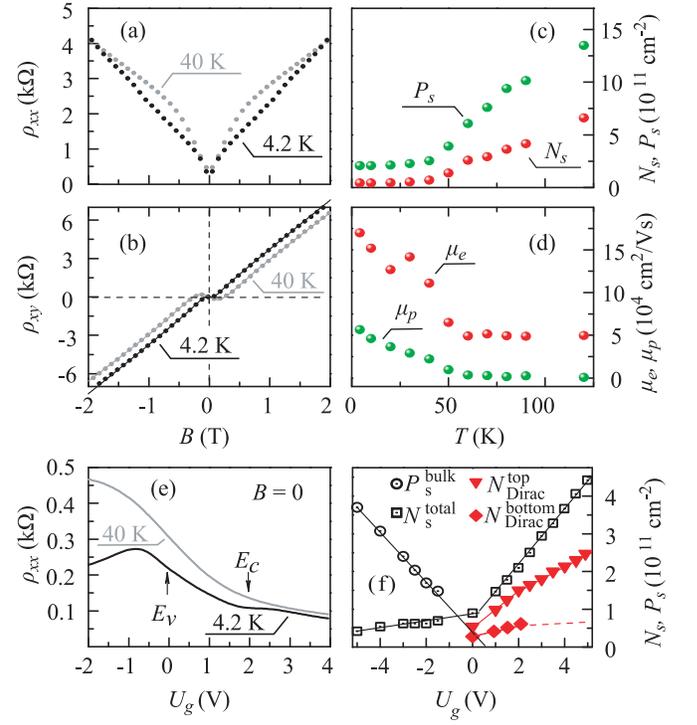}
\caption{
Magneto-transport data obtained in ungated  cross shaped (a-d),  
gated cross shaped  (e) and gated hallbar (f) samples. 
The panel (a) shows the $\rho_{xx}$ and (b) the $\rho_{xy}$ $B$-dependence for $T=4.2$\,K 
 and $T=40$\,K. (c) Electron $N_s$ and hole $P_s$ densities and (d) 
mobilities ($\mu_e$ and $\mu_p$) for different temperatures, extracted from the Drude model. 
(e) $U_g$ dependence of $\rho_{xx}$ at $T=4.2$\,K  and $T=40$\,K. The valence and conduction band edges are marked by arrows "$E_v$" and "$E_c$", respectively. 
(f) Electron and hole densities 
of bulk ($N_s^{total}$ and $P_s^{bulk})$ and 
surface states ($N_{Dirac}^{top}$ and $N_{Dirac}^{bottom}$). 
 }
\label{fig_3}
\end{figure}

Samples have been characterized by magneto-transport measurements 
using standard low-frequency lock-in technique with the 
currents in the range of $0.1-1$\,$\mu$A and with the magnetic field 
$B$ up to 7\,T applied normal to the HgTe film plane. The measurements of 
$\rho_{xx}(B)$ and $\rho_{xy}(B)$ for the ungated sample at $T =
4.2$ and 40\,K are presented in the Fig.\,\ref{fig_3}(a,b). 
The obtained $\rho_{xx}(B)$ shows a large positive magnetoresistance 
and the Hall resistance $\rho_{xy}(B)$ exhibits a nonlinear $N$-type shape around $B = 0$\,T, which both are typical for a system with coexisting electrons and holes. 
Fitting these traces by using the classical two-component Drude model 
we obtain electron and hole densities 
and mobilities~\cite{Kozlov2014}. 
The results of the fitting are presented in Fig.~\ref{fig_3}(c,d). 
The temperature dependence of electron and hole densities shows typical 
activation behavior at $T > 50$\,K 
and almost saturates below this temperature. 
The presence of holes at low temperatures indicates 
that the Fermi level is situated in the valence band
while the observed electrons are attributed to 
the surface topological states. 
Note that at high temperatures electrons populate surface as well as bulk states.
Figure~\ref{fig_3}(d) shows that both, electron and hole, 
mobilities decrease with increasing temperature. 
We emphasize that the electron mobility remains as high as 10$^5$\,cm$^2$/V$\cdot$s 
even at $T = 40$\,K.

Now we turn to magneto-transport data of gated samples. 
The $\rho_{xx}(U_g)$ dependence obtained at 
$T = 4.2$ and $40$\,K is presented in Fig.~\ref{fig_3}(e). 
The low temperature resistivity shows a maximum near $U_{g}\approx -0.8$\,V. 
At the same gate voltage the Hall resistivity 
$\rho_{xy}$ 
changes its sign at ${B = 1\,\rm{T}}$ (not shown). 
This indicates that the Fermi level can be tuned from the conduction band to the 
valence band by variation of $U_g$. 
Following the procedure described in  Ref.\,\onlinecite{Kozlov2014} we analysed
magneto-transport data and identified the gate voltages corresponding to 
the Fermi level lying at the top of the valence band and the bottom of the 
conduction band, see arrows marked "$E_{v}$" and "$E_c$"
in Fig.~\ref{fig_3}(e), respectively. 
The  electron and hole 2D density of bulk and surface states as 
a function of $U_g$ are shown in Fig.\,\ref{fig_3}(f).
This plot shows that at $U_g\lesssim 0$\,V the Fermi energy is situated in the 
valence band, i.e., holes and surface electrons coexist. At these voltages the
magnetic field dependence of longitudinal and Hall resistance for 
gated and ungated samples become similar. 
This indicates that gate fabrication does not change 
the band diagram of the structure qualitatively but only introduces a 
built-in electric field due to, e.g.,
charged impurities  and/or defects.
The defect density can be estimated as ($1 \div 2)\times 10^{11}$\,cm$^{2}$, varying from sample to sample. 

\begin{figure}[t]
\includegraphics[width=0.95\linewidth]{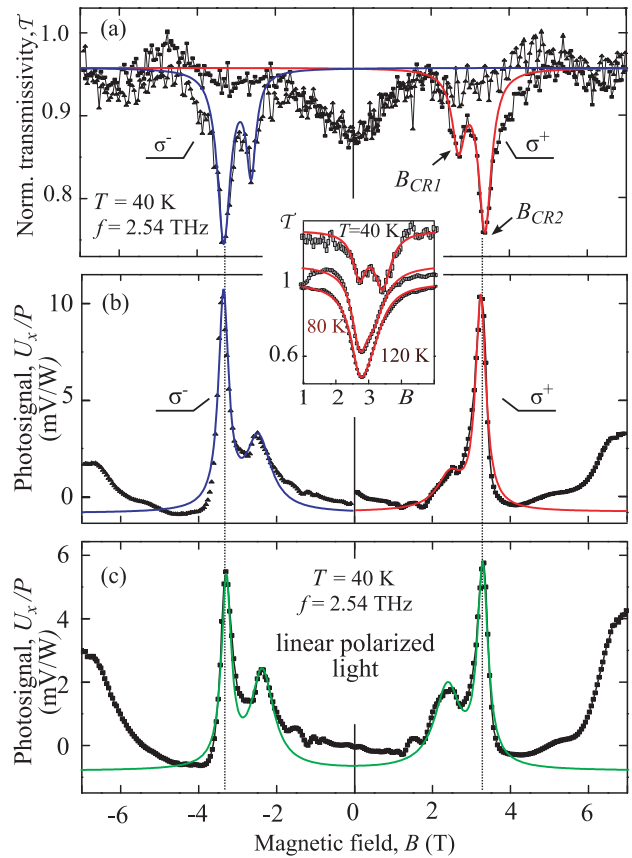}
\caption{Panel (a): Normalized transmissivity of the square shaped HgTe sample at $40$\,K as a function of magnetic field. The data are given for right ($\sigma^+$, triangles) and left ($\sigma^-$, circles) circularly polarized light. The inset shows the averaged transmissivity traces obtained for different temperatures. Panels (b) and (c) show the photosignal normalized by the radiation power ($U_x/P$) for $\sigma^+$ and $\sigma^-$ and for linear polarized radiation, respectively. Full lines in all panels show fits by the Lorentzian function.}
\label{fig_4}
\end{figure}

For photocurrent excitation and cyclotron resonance transmission measurement we apply a $cw$ molecular laser \cite{DMS2009,edge} emitting radiation with frequency $f\,$=$\,2.54$\,THz (wavelength $\lambda\,$=$\,118\,\upmu$m), $f\,$=$\,1.62$\,THz ($\lambda\,$=$\,184\,\upmu$m) and  $f\,$=$\,0.69$\,THz ($\lambda\,$=$\,432\,\upmu$m). The incident power $P\,\approx\,10$\,mW  is modulated at about 180~Hz by an optical chopper. 
Structures were placed in a temperature variable optical cryostat with $z$-cut crystal quartz windows.
The radiation at normal incidence is focused onto a spot of about 1.5\,mm diameter at the center of the sample. The spatial beam distribution has an almost Gaussian profile, measured by a pyroelectric camera~\cite{Ziemann2000}.
The initial linear polarization can be transformed into right ($\sigma^+$) and left ($\sigma^-$) handed circularly polarized light by applying a $\lambda/4$-plate. A magnetic field $B$ up to $7$\,T is applied normal to the film.
The photoresponse is measured applying standard lock-in technique, either as a photocurrent $j_{x,y}$ or the corresponding photovoltage $U_{x,y}$ picked up across $50\, \Omega$ or $10$\,M$\Omega$ load resistors, respectively. 
Parallel to the photocurrents we also studied radiation transmission, see Fig.~\ref{fig_4}(a). For that a Golay cell detector has been placed behind the sample.

\section{Experimental Results}

We begin with the transmissivity data obtained from the square shaped sample. The data for $T$\,=\,$40$\,K are shown in Fig.\,\ref{fig_4}(a) for different polarization states. Exciting the sample with right-handed circularly polarized radiation ($\sigma^+$) and sweeping the magnetic field, we observe two resonant dips  at positive magnetic fields $B$\,=\,$2.6$\,T and $B$\,=\,$3.35$\,T. Upon changing the radiation helicity from $\sigma^+$ to $\sigma^-$ the dips appear at negative magnetic fields. For linearly polarized radiation, the resonances are observed for both magnetic field directions (not shown). The resonances can be well fitted by a Lorentzian function. All these features are clear signs that the absorption of radiation is caused by cyclotron resonance of electrons. The dips in the transmissivity $\cal T$ are detected in the temperature range from $4.2$ up to $150$\,K. Below $4.2$\,K the amplitude of both dips is smaller than the noise level of our setup. With rising temperature, the two dips increase in amplitude and merge together at $T$ above $\sim 80$\,K, see inset of Fig.~\ref{fig_4}(a). The temperature dependence is due to the increasing carrier density with higher temperature and has been independently checked by magneto-transport data on the same sample, see Fig.\,\ref{fig_3}(c). 
These  results are in agreement with findings of Ref.~\onlinecite{Shuvaev, Shuvaev2013} where double CR structures have been detected in similar systems by transmission and Faraday effect applying radiation of lower frequencies and attributed to top and bottom surface states.

\begin{figure}[t]
\includegraphics[width=0.95\linewidth]{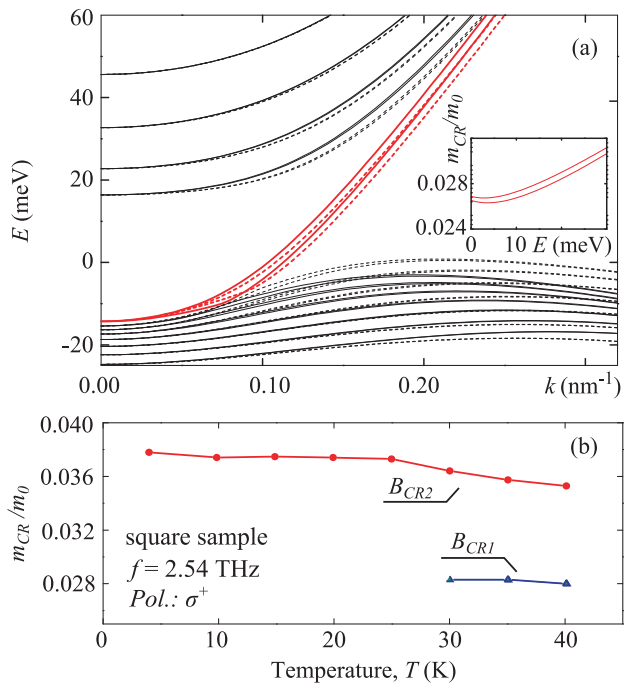}
\caption{(a) The band structure of (013)-grown $80$~nm strained HgTe film calculated by
$\bm{k}$$\cdot$$\bm{p}$ model. Solid and dashed lines show the band dispersion for two perpendicular 
in-plane crystallographic directions: $\bm k \parallel [100]$ (solid curves) and $\bm k \parallel [03\bar{1}]$ (dashed curves).
Calculation are carried out for build-in electric field $E_z = 0.5$~kV/cm which results in the energy splitting 
of the top and bottom surface states, shown by thick red solid and dashed curves. Inset shows calculated cyclotron masses for surface states as a function of energy. (b) Cyclotron masses obtained from the transmission measurements. 
Dots and triangles show the results for two CR lines.}
\label{fig_5}
\end{figure}

From the cyclotron resonance positions given by
\begin{equation}
B_{CR}=2\pi f\frac{m_{CR}}{e}
\label{eq_1}
\end{equation}
we determined the correspondent cyclotron masses, $m_{CR}$. At $T$\,=\,$40$\,K we obtained the masses 
$m_{c}$\,=\,$0.028$\,$m_{0}$ and $m_{c}$\,=\,$0.035$\,$m_{0}$, for dips at $B$\,=\,$2.6$\,T and $B$\,=\,$3.35$\,T, respectively. 
 To analyse the origin of the resonances we carried out  $\bm{k}$$\cdot$$\bm{p}$ calculations of the band structure for 
the (013)-oriented $80$\,nm wide HgTe film sandwiched between CdHgTe barriers adapting the procedure 
described in the supplementary materials to Ref.\,\onlinecite{Bruene2011} and Refs.\,\onlinecite{Zholudev2012, Novik2005}.
The details of our calculations for (013)-oriented structures taking into account build-in electric fields
are given in the Appendix. 
The obtained band structure and effective cyclotron masses for various energy levels are shown in 
Fig.\,\ref{fig_5}(a) and the inset, respectively. The calculations reveal the presence of 
surface states in the band gap of bulk strained HgTe. The Dirac points of the surface states for 
both top and bottom interfaces are deep in the valence band which leads to level mixing and deviation 
of the energy dispersion from the linear one, similarly to the results obtained for (001)-oriented 
structures\cite{Bruene2011}. 
In (013)-grown structures, the energy spectrum is anisotropic in the interface plane.  To illustrate 
the anisotropy we plotted the band structure in Fig.\,\ref{fig_5}(a) calculated for two perpendicular 
in-plane directions (solid and dashed lines). 
Moreover, compared to (001)-oriented films, 
we find that the spin carried by the surface electrons is tilted out of the interface 
plane, reflecting the low spacial symmetry of the (013)-oriented structure, see Appendix.
Our calculation shows that the masses for surface states are very close to the experimentally 
detected ones and, also in agreement with experiment, vary only weakly with temperature, see Fig.\,\ref{fig_5}(b). The detail comparison between
the experimentally observed cyclotron masses and
the calculated values require the knowledge of electric
field distribution and chemical composition profile in real structures and is out of scope of the present paper. We note that the density of bulk states in the valence band is very high due to large effective mass of holes.
Therefore, the Fermi level in ungated samples and in gated samples at negative voltages is efficiently pinned to the 
valence band top. Thus, we conclude that the resonances are caused by the top/bottom surface states. This conclusion is also supported
by magneto-transport data indicating a vanishing amount of bulk electrons. Moreover, according to the calculation, cyclotron resonance 
of the bulk carriers is expected at substantially larger magnetic fields corresponding to a mass of about $0.07$\,$m_0$.

\begin{figure}[t]
\includegraphics[width=0.95\linewidth]{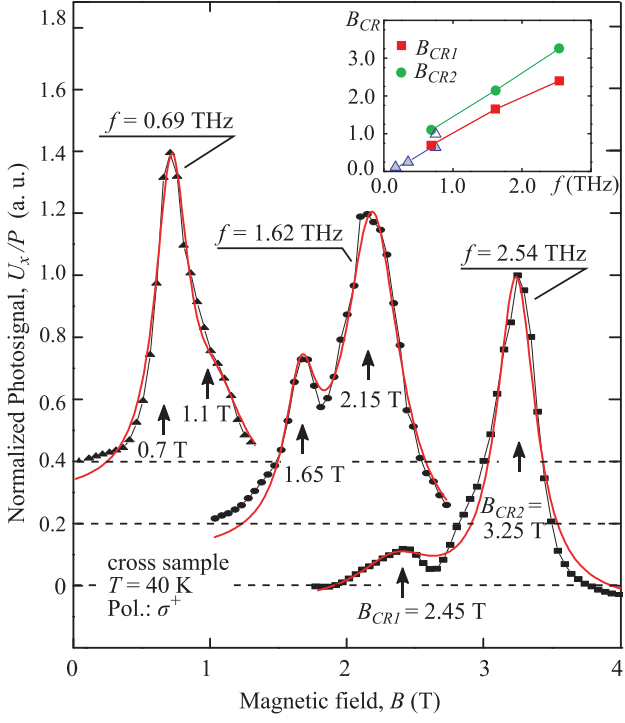}
\caption{Normalized CR radiation induced photocurrent measured for different 
radiation frequencies in ungated cross shaped samples. Note that the data are $y$-shifted 
by 0.2 for and 0.4 for $f= 1.62$ and $0.69$\,THz, respectively. Full lines  show fits by 
the Lorentzian function. Inset shows the first, $B_{CR1}$ (squares), and second, $B_{CR2}$ (circles), resonance 
positions as a function of  frequency. Full and open triangles show 
the data of Ref.\,\onlinecite{Shuvaev2012} for top and bottom surface states measured 
in similar strained HgTe film of 70~nm width.}
\label{fig_6}
\end{figure}

\begin{figure}[t]
\includegraphics[width=0.95\linewidth]{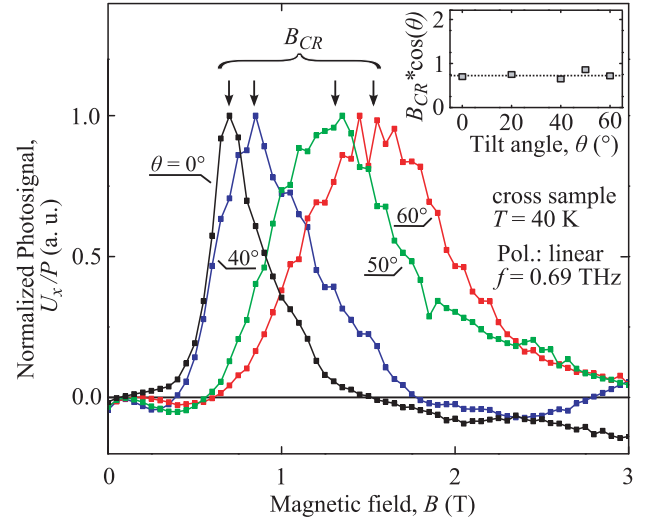}
\caption{(a) Magnetic field dependencies of the photocurrent measured for magnetic 
field tilted by an angle $\Theta$. The inset shows the resonance field,  $B^{CR}_z = B_{CR} * cos (\Theta)$, as a function of angle $\Theta$. 
}
\label{fig_7}
\end{figure}

\begin{figure}[t]
\includegraphics[width=0.95\linewidth]{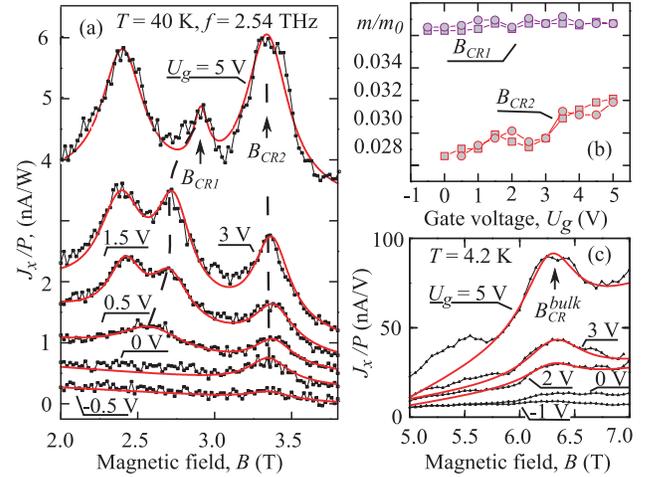}
\caption{(a) Magnetic field dependence of the photocurrent measured for gated cross sample 
$\# 1$ and different gate voltages, $U_g$. Full lines show fits by the 
Lorentzian function.
Note that the data for various $U_g$ are $y$-shifted for clearness. The shifts are: 
-0.1\,V for $U_g$ = -0.5\,V; 
0\,V for $U_g$ = 0\,V; 
0.15\,V for $U_g$ = 0.5\,V; 
0.35\,V for $U_g$ = 1.5\,V; 
0.9\,V for $U_g$ = 3\,V; 
3.5\,V for $U_g$ = 5\,V. 
(b) 
Gate voltage dependencies of effective masses determined from the position 
of CR resonances at magnetic field strength $B > 2.5$~T. 
(c) Magnetic field dependence of the photocurrent measured at 
low temperature and high magnetic fields. 
}
\label{fig_8}
\end{figure}

Next we demonstrate that the CR, so far observed in transmission, results in a photocurrent, whose magnetic field and polarization dependencies fully reproduce the peaks positions and shapes of the CR absorption. The magnetic field dependencies of the corresponding photosignal are shown for circularly or linearly polarized radiation in Fig.\,\ref{fig_4}(b) and (c), respectively. 
The coincidence of the transmission and photocurrent resonance positions indicates that the current is also excited in the surface states. For a more detailed study of the photocurrent we used cross shaped samples, which are too small to be used for transmission measurements, but allow fabrication of semitransparent gates. Figure 
\ref{fig_6} shows the resonant photocurrent obtained for different radiation frequencies in an ungated cross shaped sample. Comparison of the data for $f = 2.54$\,THz obtained for square and crossed shaped samples, shows that magnetic field positions of the resonances stay almost the same. Notice that the relative strength of the first, $B_{CR1}$, and second, $B_{CR2}$, resonances substantially change with decreasing frequency. While at $f = 2.54$\,THz the second resonance is substantially stronger than the first one, for $f = 0.69$\,THz the situation reverses. 
The inset in Fig.\,\ref{fig_6} demonstrates that the position of both resonances linearly scales with the frequency. The frequency dependence of the CR is in line with transmissivity data of Ref.\,\onlinecite{Shuvaev2012} of top and bottom surface states, obtained for a strained HgTe films of 70~nm width, see triangles in the inset in Fig.\,\ref{fig_6}.
In order to provide additional support for the conclusion that the observed resonant photocurrent and CR stem from two-dimensional surface states, we carried out measurements with the magnetic field tilted by an angle $\Theta$. Figure\,\ref{fig_7} shows that the current is caused by the magnetic field component $B_z$, normal to the surface. Indeed, calculating from the peak position $B_{CR}$ in Fig.\,\ref{fig_7}, the value of $B^{CR}_z = B_{CR} * cos (\Theta)$, we obtain that $B^{CR}_z$ does not depend on the angle $\Theta$, see inset. Moreover, for an in-plane magnetic field the photocurrent vanishes (not shown), indicating the two-dimensional nature of the carriers~\cite{footnote1}.

\begin{figure}[t]
\includegraphics[width=0.95\linewidth]{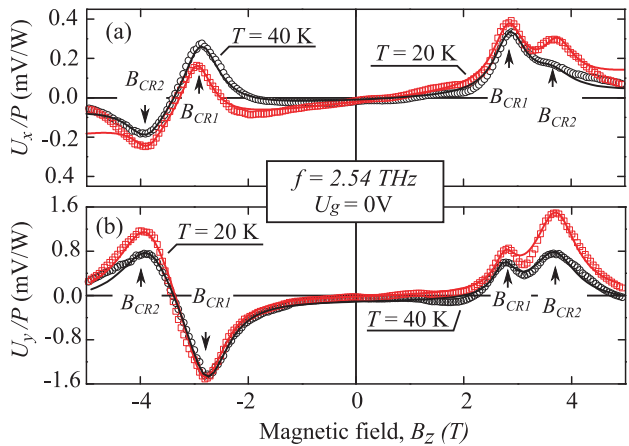}
\caption{Magnetic field dependence of the photosignal measured for  
two in-plane directions ($x$ and $y$) obtained for gated cross sample $\# 2$
($U_g = 0 $). }
\label{fig_9}
\end{figure}

By using gated cross shaped samples we study the photocurrent and CR as a function of the Fermi energy, see Fig.\,\ref{fig_8}(a). The data show that at negative bias voltages, for which the Fermi energy lies in the valance band, the photocurrent resonances vanish. For zero gate voltage a resonance at $B = 3.3$\,T is clearly resolved.  With increasing gate voltage we additionally detected a resonance at smaller magnetic field, see data for $U_g = 0.5$\,V in Fig.\,\ref{fig_8}(a). The positions of the resonances are similar to the ones obtained for ungated samples, see Figs.\,\ref{fig_4} and \ref{fig_6}, and correspond to the response of the top and the bottom surface states of the film. At $U_g > 1$\,V we detect an additional resonance at $B \approx 2.4$\, T. Note that for $U_g > 1$\,V and $T=40$\,K the Fermi energy lies close to the bottom of the conduction band resulting in a noticeable electron density in the bulk, see Fig.\,\ref{fig_3}(e). Moreover, the high gate voltage applied to the film may form an inversion layer
for bulk electrons changing their dispersion. 
The origin of this additional resonance is out of scope of the present paper.
%
At even higher gate voltages $U_g > 2$\,V and  $T= 4.2$\,K we detected another resonance at much stronger magnetic fields $B\approx 6.3$\,T, see Fig.\,\ref{fig_8}(c). The position of this resonance does not depend on the gate voltage and the corresponding effective mass is $m = 0.07$\,$m_0$. This mass matches  the one calculated for bulk electrons, so that the resonance can be attributed to the CR of bulk electrons. We note that at $T = 4.2$\,K the surface state resonances are superimposed by Shubnikov-de Haas oscillations, which makes the CR analysis almost impossible.  

From the resonance positions of the surface state response we determine the gate voltage dependence of the effective masses, see Fig.\,\ref{fig_8}(b). 
While the mass corresponding to $B_{CR1}$ remains almost independent of the gate voltage, the $B_{CR2}$ slightly rises 
with increasing $U_g$. 
The resonance at $B_c \approx 3.3$\,T ($m_{CR} \approx 0.036 m_0$) is attributet to the bottom surface, whose electron density is not much influenced by the gate voltage.
In fact it is known from the magneto-transport measurements, see Fig.\,\ref{fig_3}(f), that the filling rates $dN_s/dU_g$ for top and bottom surface states are considerable different because of electrostatic screening of the bottom surface by the top one.
Moreover the electron density at the top surface rises with the gate voltage increase, see Fig.\,\ref{fig_3}(f), which leads to the increase of the cyclotron mass. By contrast the density of the bottom electrones is almost independent on the gate voltages, see Fig.\,\ref{fig_3}(f).
From the full widths at half maximum of the CR measured in transmission in square shaped ungated samples, see Fig.\,\ref{fig_4}(a), we also estimate the mobility of the surface electrons, being $\mu =5.5 \cdot 10^4$\,cm$^2$/V$\cdot$s.
This value fits well to the mobility obtained from magneto-transport in ungated cross shaped samples, see Fig.\,\ref{fig_3}(d). 
Finally, we discuss the photocurrent behavior when reversing the magnetic 
field direction. The magnetic field dependence of the photocurrent measured for  
two in-plane directions ($x$ and $y$) is shown in Fig.\,\ref{fig_9}. 
First of all this plot indicates that both resonances behave independently 
of each other. Indeed, while the resonance for the photosignal $U_x$ at $B_{CR1}$ 
is even in magnetic field, the one at $B_{CR2}$ is odd. This result agrees well 
with the above conclusion, that the photocurrents are generated by two independent 
surface states (top and bottom). For the other direction this situation just reverses: 
Now the resonance at $B_{CR1}$ is odd and $B_{CR2}$ is even in magnetic field.

This $B$-dependence of the photocurrent can be understood 
from the phenomenological description based on symmetry arguments. 
The appearance of magnetic field induced photocurrents, excited by
homogeneous radiation, is well known for 
two dimensional systems lacking  spatial inversion 
symmetry, for review see Ref.\,\onlinecite{BelkovSST,Belkovbuch}.
For free carrier absorption, being accompanied by electron scattering 
from phonons or static defects, the current is caused by 
the asymmetry of electron scattering in 
momentum space induced by an external magnetic field, when the electron 
systems is driven out of thermal equilibrium by light absorption \cite{Tarasenko2008}. 
At cyclotron resonance, the electron-photon interaction is enhanced, which 
leads to a resonant behavior of the photocurrent. 

Considering the generation of asymmetric electron distribution in the momentum space by $cw$ radiation,
the distribution evolution in the magnetic field and its decay due to scattering processes, one can write
the Boltzmann equation for the distribution function $f_{\bm{p}}$ 
\begin{equation}
\label{kinetic_equation}
e [\bm{v} \times \bm{B}] \cdot \dfrac{\partial f_{\bm{p}}}{\partial \bm{p}} = g_{\bm{p}}-\dfrac{f_{\bm{p}}-\langle f_{\bm{p}} \rangle}{\tau}\:,
\end{equation}
where $\bm{p}$ and $\bm{v}$ are the electron momentum and velocity, respectively, $g_{\bm{p}}$ is the generation rate of electrons in the state with the momentum $\bm{p}$, $\tau$ is the momentum relaxation time, and $\langle f_{\bm{p}} \rangle$ is the distrubution function averaged over the momentum direction. Multiplying Eq.~\eqref{kinetic_equation} by $e \bm{v}$ and summing up over $\bm{p}$ we obtain
the equation for the photocurrent density $\bm{j} = \sum_{\bm{p}} e \bm{v} f_{\bm{p}}$ in the absence of an in-plane bias
\begin{equation}\label{phenom}
\bm{j} \times \bm{\omega}_c = \bm{G} - \frac{\bm{j}}{\tau} \:,
\end{equation}   
where $\bm{\omega}_c = \omega_c \hat{\bm{z}}$, $\omega_c = e B_z /m_{CR}$ is the cyclotron frequency, $\hat{\bm{z}}$ is the unit vector 
pointing along the $z$ axis, and $\bm{G}=\sum_{\bm{p}} e \bm{v} g_{\bm{p}}$ is the rate of current generation.

\begin{figure}[t]
\includegraphics[width=0.95\linewidth]{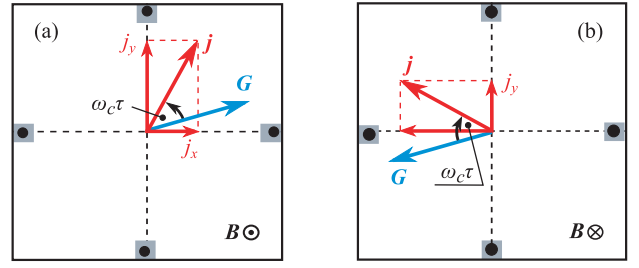}
\caption{The vectors of photocurrent generation rate $\bm{G}$ and the steady-state photocurrents 
$\bm{j}$ for two opposite magnetic field directions, panels (a) and (b). The direction of photocurrent $\bm{j}$ in the magnetic field 
is declined from the direction of $\bm{G}$ by the Hall angle $\arctan (\omega_c \tau)$. Figure illustrates that, for the given parameters, $j_x$ changes its sign while $j_y$ converses its sign when the magnetic field direction is reversed. 
}
\label{fig_10}
\end{figure}

The direction of the vector $\bm{G}$ is determined by the point group symmetry of the structure, 
radiation polarization state and magnetic field orientation. 
In topological states on (013)-oriented surfaces with no nontrivial symmetry elements (C$_1$ point group), 
the current generation becomes possible even at normal incidence of radiation and
magnetic field perpendicular to the surface. In this particular geometry
the generation rate $\bm{G}$ can be presented in the form
\begin{equation}
\bm{G} = \bm{\gamma} I \eta(\omega) B_z\:,
\end{equation}

where $I$ is the radiation intensity, $\eta(\omega)$ is absorbance and 
$\bm{\gamma}$ is the vector determined by the magnetic field induced asymmetry of 
electron scattering processes. 
%
%
For C$_1$ point group the direction of the vector $\bm{\gamma}$
is not forced to a certain crystallographic axis by symmetry arguments 
and may depend on temperature, gate voltage, radiation frequency, etc. 
Such a specific anisotropy has been demonstrated for photogalvanic currents 
excited by terahertz radiation in (013) HgTe quantum wells 
at zero magnetic field~\cite{Wittman2010}.
For two perpendicular in-plane directions the solution of Eq.~\eqref{phenom}  has the form 
\begin{eqnarray}\label{fin_eq}
j_{x} = \frac{\gamma_x - \omega_c \tau \, \gamma_y }{1+(\omega_c \tau)^2} I \eta(\omega) B_z\:, \\
j_{y} = \frac{\gamma_y + \omega_c \tau \, \gamma_x }{1+(\omega_c \tau)^2} I \eta(\omega) B_z\:.\nonumber
\end{eqnarray}   
Equation~\eqref{fin_eq} shows that the current (i) has a resonant behavior at $\omega=\omega_c$ caused by CR in the absorbance $\eta(\omega)$ and (ii) contains contributions even and odd in the magnetic field, both in agreement with experiment.The formation of photocurrent for positive and negative magnetic fields $B_z$ is illustrated in Fig.~\ref{fig_10}.
sketches the generation rate vector $\bm{G}$ for positive magnetic fields. The resulting photocurrent $\bm{j}$
deviates from $\bm{G}$ due to the Hall effect. For opposite magnetic field 
the vector $\bm{G}$ and the Hall rotation axis reverse as shown in Fig.~\ref{fig_10}(b). Consequently, for $x$ and $y$
projections we obtain both odd and even components of the total photocurrent, see Fig.~\ref{fig_9}

\section{Summary}

To summarize, our results demonstrate that excitation of 
80~nm thick strained HgTe films by terahertz radiation results 
in cyclotron resonance induced photocurrents generated in the 
topologically protected surface states. The effect emerges due to 
magnetic field induced asymmetric scattering of 
nonequilibrium surface carriers in the momentum space. 
Our study, accompanied by complimentary transmission and magneto-transport measurements
show that the observed CR assisted photocurrent provides a sensitive method to probe the 
effective masses of 2D Dirac surface states. Importantly, the method is applicable even for 
small size samples, e.g. gated structures, and for the Fermi energy lying in the conduction band, i.e.\,conditions where transmission experiments are almost impossible.

\section{Appendix}

We calculate electron spectrum in low-symmetry strained CdHgTe/\-HgTe/CdHgTe structure on CdTe 
in the framework of 6-band $\bm{k}$$\cdot$$\bm{p}$ model which is relevant for narrow gap
semiconductors. We consider the class of $(0 l h)$-oriented films ($l$ and $h$ are integer numbers) 
which includes (011)-, (012)-, and (013)-oriented structures. In the basis of Bloch amplitudes of the $\Gamma_6$ and $\Gamma_8$ bands, the wave function has the form
\begin{equation}
\Psi(\bm{\rho},z) = \left( \begin{array}{c} \psi_{\Gamma_6, +1/2} \\ \psi_{\Gamma_6, -1/2} \\
\psi_{\Gamma_8, +3/2} \\ \psi_{\Gamma_8, +1/2} \\
\psi_{\Gamma_8, -1/2} \\ \psi_{\Gamma_8, -3/2} \end{array} \right) \exp(i \bm{k}_{\parallel} \cdot \bm{\rho}) ,
\end{equation}
where $\bm{\rho}=(x,y)$ is the in-plane coordinate, $z$ is the growth axis, $\psi_j$ are the envelopes, $\bm{k}_{\parallel}=(k_x,k_y)$ is the in-plane wave vector. 

\begin{table}[b]\label{table1}
\begin{flushleft}
  \begin{tabular*}{0.46\textwidth}{ l | l  l  l  l  l  l}
    \hline
           & $E_g $ (eV)     \;   & $\gamma_1$   \;       & $\gamma_2$     \;     & $\gamma_3$  \;   & $2 m_0 (P/\hbar)^2 $ (eV) \\
    \hline
    HgTe\: & -0.303$^{(1)}$    & 4.1$^{(1)}$        & 0.5$^{(1)}$        & 1.3$^{(1)}$   &  18.8$^{(1)}$     \\ 
    CdTe\: & 1.606$^{(1)}$     & 1.47$^{(1)}$       & -0.28$^{(1)}$      & 0.03$^{(1)}$  &  18.8$^{(1)}$   \\
    \hline
   \end{tabular*}
   \begin{tabular*}{0.46\textwidth}{ l | l  l  l  l  l  l}   
    \hline
           & $F$     \;       & $a$ (eV)      \;       & $b$ (eV)   \;   &  $d$ (eV)  \;   & $\Xi_c$ (eV) \\
    \hline
    HgTe\: & 0$^{(1)}$     & -0.13$^{(2)}$  & -1.5$^{(4)}$  & -8.0$^{(4)}$  & -3.82$^{(6)}$  \\
    CdTe\: & -0.09$^{(1)}$ & 0.756$^{(3)}$        & -1.0$^{(5)}$ & -4.4$^{(5)}$ & -2.687$^{(3)}$  \\    
    \hline
   \end{tabular*}
   \begin{tabular*}{0.46\textwidth}{ l | l  l  l  l  l  l}   
    \hline
             & $a_0$ (\AA)  \;  & $c_{11}$ (Mbar)  \,   & $c_{12}$ (Mbar)   \,  & $c_{44}$ (Mbar)\\
    \hline
    HgTe\:   & 6.46$^{(5)}$  & 0.597$^{(2)}$ & 0.415$^{(2)}$ & 0.226$^{(2)}$\\
    CdTe\:   & 6.48$^{(5)}$  & 0.562$^{(2)}$ & 0.394$^{(2)}$ & 0.206$^{(2)}$\\
    \hline
  \end{tabular*}
\end{flushleft}
\caption{Parameters of HgTe and CdTe. $^{(1)}$Ref.~\onlinecite{Novik2005}, $^{(2)}$Ref.~\onlinecite{Chris1989},
$^{(3)}$Ref.~\onlinecite{Merad2003}, $^{(4)}$Ref.~\onlinecite{Yakunin1988}, $^{(5)}$Ref.~\onlinecite{Adachi2004}, 
$^{(6)}$Ref.~\onlinecite{Latussek2005}.}
\end{table}

The effective 6-band Hamiltonian is given by
\begin{equation}\label{Hamiltonian}
H = \left( 
\begin{array}{cc}
H_{cc} & H_{cv} \\
H_{cv}^\dag & H_{vv}
\end{array}
\right) ,
\end{equation}
where the blocks $H_{cc}$ and $H_{vv}$ describe the conduction ($\Gamma_6$)  and valence ($\Gamma_8$) intraband contributions, and the block $H_{cv}$ describes the band mixing. The block $H_{cc}$ is given by
\begin{equation}
H_{cc} = I_{2\times2} \left[E_{c}(z) + \frac{\hbar^2 \, \bm{k} [2F(z)+1] \bm{k}}{2m_0}  + \Xi_c \, {\rm Tr} \epsilon \right] \:,
\end{equation}
where $I_{2\times2}$ is the $2\times2$ identity matrix, $E_c(z)$ is the conduction band profile, $\bm{k}=(k_x,k_y, -i \partial/\partial_z)$, $F(z)$ is a parameter accounting for contribution from remote bands, $\Xi_c$ is the $\Gamma_6$-band deformation potential constant, and $\epsilon$ is the train tensor.
The block $H_{cv}$ has the form 
\begin{equation}
H_{cv} = \left( 
\begin{array}{cccc}
-\dfrac{1}{\sqrt{2}} P k_{+} & \sqrt{\dfrac{2}{3}}P k_z & \dfrac{1}{\sqrt{6}} P k_-& 0 \\
0 & -\dfrac{1}{\sqrt{6}} P k_+ & \sqrt{\dfrac{2}{3}} P k_z & \dfrac{1}{\sqrt{2}} P k_+
\end{array}
\right) ,
\end{equation}
where $P$ is the Kane matrix element and $k_{\pm}=k_x\pm i k_y$.

The block $H_{vv}$ is given by
\begin{equation}
H_{vv} = E_v(\bm{r}) + H_{L}^{({\rm i})} + H_{L}^{({\rm a})} + H_{BP}^{({\rm i})} + H_{BP}^{({\rm a})}\:,
\end{equation}
where $E_v(\bm{r})$ is the valence band profile, $H_{L}^{({\rm i})}$, $H_{L}^{({\rm a})}$, $H_{BP}^{({\rm i})}$, and $H_{BP}^{({\rm a})}$
are the isotropic and anisotropic parts of the Luttinger and Bir-Pikus Hamiltonians,  
\begin{equation}
H_{L}^{({\rm i})} = \frac{\hbar^2}{2 m_0} \left[ -{\bm k} \left( \gamma_1 + \frac52 \gamma_2\right) {\bm k} + 2 (\bm{J}\cdot \bm{k}) \gamma_2 (\bm{J}\cdot \bm{k}) \right]\:,
\end{equation}
\begin{equation}
H_{BP}^{({\rm i})} = \left(a+\dfrac{5}{4} b\right) {\rm Tr} \epsilon - b \sum_{\alpha} J_{\alpha}^2 \, \epsilon_{\alpha\alpha} 
- b \sum_{\alpha \neq \beta} \{ J_{\alpha}, J_{\beta}\}_s \epsilon_{\alpha\beta}  \:,
\end{equation}
$\gamma_1$, $\gamma_2$ and $\gamma_3$ are contributions to the Luttinger parameters from remote bands, $\bm{J}$ is the vector composed of the matrices of the angular momentum $3/2$, $\{J_\alpha, J_\beta \}_s = (J_\alpha J_\beta  + J_\beta J_\alpha )/2$, $a$, $b$ and $d$ are the $\Gamma_8$-band deformation potential constants.
The explicit form of the terms $H_{L}^{({\rm a})}$ and $H_{BP}^{({\rm a})}$ related to the cubic anisotropy of host crystals depends on the coordinate frame used. 

In the coordinate frame relevant to $(0lh)$-oriented structures: $x \parallel [100]$, $y \parallel [0 h\overline{l}]$, and $z \parallel [0 l h] $, the terms $H_{L}^{({\rm a})}$ and $H_{BP}^{({\rm a})}$ assume the form
\begin{widetext}
\begin{equation}
H_{L}^{({\rm a})} = \frac{2\hbar^2}{m_0} \left\{ \{J_x J_y\}_s  
(\gamma_3-\gamma_2)k_x k_y + \{J_x J_z\}_s \{\gamma_3-\gamma_2, k_z \}_s\, k_x + \left[\{J_y J_z\}_s \cos 2\theta + \frac{J_z^2 - J_y^2}{2} \sin 2\theta \right] \right.
\end{equation}
\[
\times \left. \left[\{(\gamma_3-\gamma_2),k_z\}_s \, k_y \cos 2\theta + \frac{k_z (\gamma_3-\gamma_2) k_z  - (\gamma_3-\gamma_2)k_y^2}{2} \sin 2\theta \right] \right\}\:,
\]
\[
H_{BP}^{({\rm a})} = - 2 \left( \frac{d}{\sqrt{3}} -b \right) \left\{ \{J_x J_y\}_s \epsilon_{xy} + \{J_x J_z\}_s \epsilon_{xz}  + \left[\{J_y J_z\}_s \cos 2\theta + \frac{J_z^2-J_y^2}{2} \sin 2\theta \right] \left[\epsilon_{yz} \cos 2\theta + \frac{\epsilon_{zz}-\epsilon_{yy}}{2} \sin 2\theta \right] \right\},
\]
\end{widetext}
where $\theta = \arctan (l/h)$ is angle between the growth direction $[0 l h]$ and the $[001]$ axis, $\theta 
 \approx 18.4^\circ$ for (013)-oriented structures.

HgTe and Cd$_{0.65}$Hg$_{0.35}$Te layers adopt the in-plane lattice structure of CdTe buffer which leads to the 
in-plane strain of the layers described by the strain tensor components $\epsilon_{xx}=\epsilon_{yy}=a_{{\rm CdTe}}/a_0-1$, where
$a_{{\rm CdTe}}$ is the lattice constant of CdTe and $a_0$ is the equilibrium lattice constant of the considered layer. Out-of-plane components of the strain tensor in each layer can be found by minimizing the elastic energy. Such calculations for $(0 l h)$-oriented structures yield
\begin{widetext}
\begin{equation}
\epsilon_{zz} = \dfrac{ c_{11}^2+2 c_{11} (c_{12}-c_{44})+c_{12}(-3 c_{12}+10 c_{44})-(c_{11}+3 c_{12}) (c_{11}-c_{12}-2 c_{44}) \cos 4\theta }{-c_{11}^2-6 c_{11} c_{44}+c_{12} (c_{12}+2 c_{44})+(c_{11}+c_{12}) (c_{11}-c_{12}-2 c_{44}) \cos 4 \theta } \epsilon_{\|} \:,
\end{equation}
\[
\epsilon_{zy} = \dfrac{ (c_{11}+2 c_{12}) (c_{11}-c_{12}-2 c_{44}) \sin 4 \theta}{-c_{11}^2-6 c_{11} c_{44}+c_{12} (c_{12}+2 c_{44})+(c_{11}+c_{12}) (c_{11}-c_{12}-2 c_{44}) \cos 4 \theta } \epsilon_{\|} \:,
\]
\end{widetext}
$\epsilon_{xz}=0$, where $c_{11}$, $c_{12}$, and $ c_{44}$ are elastic constants. 

\begin{figure}[b]
\includegraphics[width=0.9\linewidth]{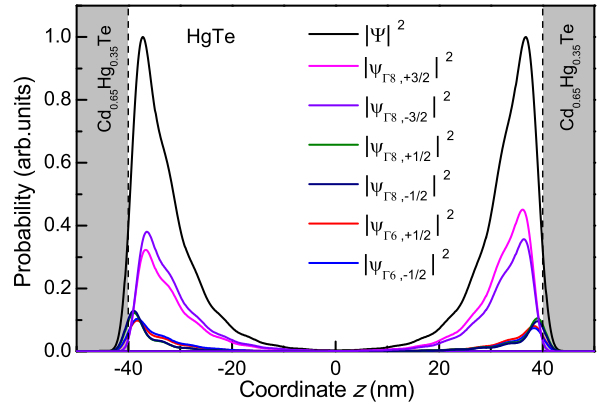}
\caption{Wave functions of edge states in (013)-grown $80$~nm strained HgTe film. Calculations are carried out for the electron 
energy $E = 11$~meV, the wave vector components $k_x > 0$ and $k_y = 0$,  and the build-in electric field $E_z = 2$~kV/cm.}
\label{figure_11}
\end{figure}

We have performed calculation of the electron spectrum and wave functions for the parameters listed in Table~I. The 
Sch\"{o}dinger equation with the matrix Hamiltonian~\eqref{Hamiltonian} is solved numerically by expanding the envelope functions 
$\psi_j(z)$ in series of the harmonic oscillator functions following the procedure described for (001)-oriented structures in Ref.\,\onlinecite{Novik2005}. The calculated spectrum for the (013)-oriented structure is shown in Fig.~\ref{fig_5}. 
The wave functions of states emerging in the band gap of bulk strained HgTe
are shown in Fig.~\ref{figure_11}. The states are localized at the film interfaces and predominantly formed from the heavy-hole states.
Despite the fact that 6-band $\bm{k}$$\cdot$$\bm{p}$ model used in the calculations neglects bulk inversion asymmetry of HgTe and the corresponding Dresselhaus spin-orbit interaction, we have found that the spin carried by surface states in (013)-oriented structures for $k_x \neq 0$ has an out-of-plane component. The out-of-plane polarization defined by
\begin{equation}
P_z = \frac{\int (|\psi_{\Gamma_8,+3/2}|^2 - |\psi_{\Gamma_8,-3/2}|^2 ) \,dz}
{\int (|\psi_{\Gamma_8,+3/2}|^2 + |\psi_{\Gamma_8,-3/2}|^2 ) \, dz}
\end{equation}
is about $0.1$ for the surface states presented in Fig.~\ref{figure_11}. The emergence of the out-of-plane spin component is attributed to the cubic anisotropy incorporated in the Hamiltonian~\eqref{Hamiltonian} together with inversion symmetry breaking at surfaces. Indeed, the 6-band $\bm{k}$$\cdot$$\bm{p}$ Hamiltonian of bulk crystal corresponds to the $O_h$ point group containing the spatial inversion. At a surface, the symmetry $z \leftrightarrow -z$ is broken which leads, for (013)-oriented surface, to the reduction of system symmetry to the $C_s$ point group containing the only non-trivial element $x \leftrightarrow -x$. Such a symmetry reduction enables the emergence of out-of-plane spin polarization for surface states with $k_x \neq 0$. We note that Dresselhaus spin-orbit interaction steming from bulk inversion asymmetry of the host zinc-blende structure, which in not included in the model Hamiltonian, will lead to an out-of-plane spin polarization for surface states with $k_y \neq 0$.

\section{Acknowledgments}

The authors thank E.G. Novik for discussions. The work was supported by DFG (SPP 1666) and RFBR.
K.-M.D., P.O., D.W. and S.D.G.\,acknowledges support from Elite Network of Bavaria (K-NW-2013-247). G.V.B.\,and S.A.T.\,acknowledge support from
RF President Grant No. MD-3098.2014.2 and the ``Dynasty'' Foundation.


\begin{thebibliography}{99}
\bibitem{footnote1}
Note, that the corresponding magnetic length  $l_{B}=\sqrt{\dfrac{\hbar}{e B}}$ being about $16$\,nm for $B\sim 3.3$\,T is smaller than the thickness of HgTe films ($80$\,nm), so that for the bulk carriers one would expect CR even for in-plane magnetic fields.
\end{thebibliography}
\end{document}